\documentclass[12pt]{article}%
\usepackage{amsfonts}
\usepackage{amsmath,amsthm,amssymb}
\usepackage{epic, eepic}

\usepackage{amsmath}
\usepackage{amssymb}%
\providecommand{\U}[1]{\protect\rule{.1in}{.1in}}
\newtheorem{theo}{Theorem}[section]

\newtheorem{exam}{Example}[section]
\newtheorem{defi}{Definition}[section]
\newtheorem{rema}{Remark}[section]

\newtheorem{maco}{Matching Condition}
\begin{document}

\title{Large Deviation Strategy for Inverse Problem}
\author{Izumi Ojima and Kazuya Okamura\vspace*{3mm}\\{\normalsize Research Institute for Mathematical Sciences, Kyoto University,}\\{\normalsize Kyoto 606-8502, Japan}}
\date{\today}
\maketitle

\begin{abstract}
Taken traditionally as a no-go theorem against the theorization of inductive
processes, Duhem-Quine thesis may interfere with the essence of statistical
inference. This difficulty can be resolved by \textquotedblleft Micro-Macro
duality\textquotedblright\ \cite{Oj03, Oj05} which clarifies the importance of
specifying the pertinent aspects and accuracy relevant to concrete contexts of
scientific discussions and which ensures the matching between what to be
described and what to describe in the form of the validity of duality
relations. This consolidates the foundations of the inverse problem, induction
method, and statistical inference crucial for the sound relations between
theory and experiments. To achieve the purpose, we propose here Large
Deviation Strategy (LDS for short) on the basis of Micro-Macro duality,
quadrality scheme, and large deviation principle. According to the quadrality
scheme emphasizing the basic roles played by the dynamics, algebra of
observables together with its representations and universal notion of
classifying space, LDS consists of four levels and we discuss its first and
second levels in detail, aiming at establishing statistical inference
concerning observables and states. By efficient use of the central measure, we
will establish a quantum version of Sanov's theorem, the Bayesian escort
predictive state and the widely applicable information criteria for quantum
states in LDS second level. Finally, these results are reexamined in the
context of quantum estimation theory, and organized as quantum model
selection, i.e., a quantum version of model selection.
\end{abstract}

\section{Statistical Inference vs. Duhem-Quine Thesis}

The main purpose of the present paper is to propose a general method for
statistical inference which we call Large Deviation Strategy (LDS for short).
To see the importance of this task, we first contrast it with the following
famous dilemma of Duhem-Quine thesis. \newline\textbf{Duhem-Quine thesis:}
It is impossible to determine uniquely such a theory from phenomenological
data as to reproduce the latter, because of unavoidable finiteness in number
of measurable quantities and of their limited accuracy.

According to the standard interpretation of this thesis as a no-go theorem
against the possibility of theorizing inductive processes, the communities of
sciences (and philosophy of sciences) have long been dominated by such common
and/or implicit consensus that the inductive aspects can be treated only in
intuitive and heuristic manners without being incorporated into theories where
only deductive arguments can be developed from some tentative and ad hoc
starting postulates without satisfactory bases. In this situation, we would
totally lose any sound basis for the mutual connections between experimental
and theoretical sides, by which any attempts for statistical estimates and
inference would become meaningless. While this issue is seldom taken serious
by working scientists such as physicists, the reason still remains to be
explained why experimental sciences can work in spite of this no-go theorem;
this question cannot be answered by the present-day forms of sciences (nor by
philosophy of sciences) in the standard formulation, for lack of the
theoretical elements of induction.  Since \textquotedblleft Macro\textquotedblright\ from the standard
viewpoint of microscopic physics is nothing more than rough approximations of
\textquotedblleft Micro\textquotedblright\ levels, such important theoretical
roles played by it as universal reference systems or its origin are hardly
examined, and hence, no justification can be given of the status of
\textquotedblleft Macro\textquotedblright. In the light of the above
Duhem-Quine thesis, therefore, it becomes evident not only that the sacred
\textquotedblleft Micro\textquotedblright\ theory itself in the usual
approaches is just something postulated in an \textit{ad hoc} way without any
inevitable basis, for lack of the unique choices of theoretical starting
points on the \textquotedblleft Micro\textquotedblright\ side in relation to
the \textquotedblleft Macro\textquotedblright\ data, but also that the
latter side is floating in the air without firm bases.

In sharp contrast, the formulation based on \textquotedblleft Micro-Macro
duality\textquotedblright\ \cite{Oj05}\ proposed by one of the authors (I.O.)
resolves the above conflict in a natural way, on the basis of the duality
between the \textquotedblleft Micro\textquotedblright\ side to be described
and the \textquotedblleft Macro\textquotedblright\ side to describe.
Therefore, it is necessary for the essence of \textquotedblleft Micro-Macro
duality\textquotedblright\ to be discussed .

\subsection{Micro-Macro duality solving Duhem-Quine thesis\newline and
quadrality scheme}

The notion of dualities can be formulated mathematically in its general form
as categorical adjunctions \cite{MacL} materializing the important aspects of
mathematical universalities. In this context, \textquotedblleft
Micro\textquotedblright\ and \textquotedblleft Macro\textquotedblright\ are
interrelated with each other by \textquotedblleft Micro-Macro
duality\textquotedblright\ in bi-directional ways: \textquotedblleft
Macro\textquotedblright\ playing the roles of a standard reference frame is
generated as a stabilized domain through the processes of emergence
\cite{Oj102} from the dynamical motions in \textquotedblleft
Micro\textquotedblright. In the opposite direction, \textquotedblleft
Macro\textquotedblright$\Longrightarrow$ \textquotedblleft
Micro\textquotedblright, the extended machinery based on the so-called
\textquotedblleft dilation\textquotedblright\ method allows us to recover the
original microscopic system, \textquotedblleft Micro\textquotedblright, from
phenomenological and/or experimental data in \textquotedblleft
Macro\textquotedblright, by means of such generalizations of the inverse
Fourier transform as Tannaka-Krein-Tatsuuma duality \cite{EnckSchw,Krein,Tann,Tatsu} and as Galois
extensions materialized by crossed product formation \cite{T02}. In this way,
the essence of the Micro-Macro duality can be understood as the adaptations to
natural sciences of the mathematical notion of duality (or adjunction)
appearing ubiquitously in mathematics.

What is most important here is the validity of mathematical universalities,
which resolves the difficulties caused by the no-go theorem of Duhem-Quine
thesis in the following way. According to the thesis, we cannot avoid any kind of
indeterminacy on the phenomenological \textquotedblleft
Macro\textquotedblright\ side\ based on the statistical inference, because of
the inevitable finiteness in number of measurable quantities and of their
limited accuracy, which will lead to possible non-uniqueness of the results of
inductions in the form of a theoretical starting point of \textquotedblleft
Micro\textquotedblright\ extracted from the phenomenological \textquotedblleft
Macro\textquotedblright. Because of the \textit{universality} associated to
\textquotedblleft Micro-Macro duality\textquotedblright, the duality between
\textquotedblleft Micro\textquotedblright\ and \textquotedblleft
Macro\textquotedblright, the uniqueness of \textquotedblleft
Micro\textquotedblright\ is guaranteed \textit{within the context} specified
by the \textquotedblleft Macro\textquotedblright\ in such forms as the
relevant aspects and accuracies compatible with the phenomenological data.

In the standard approach in physics concentrating on the unilateral efforts to
derive experimental predictions from theoretical hypotheses on the purely
\textquotedblleft Micro\textquotedblright\ side, this kind of approach might
be unfamiliar. So, we try to explain briefly the essence of some key notions
relevant to duality. First, the notion of duality is widely applied in many
mathematical contexts in such a form as the duality between an abstract group
and (the totality of) its representations. One can also find duality in
physics in such a form as position $x$ and momentum $p$ as it is essential for
the basis of many concepts. The most typical example in the present context is
the duality between observables and states. In the algebraic formulation of
quantum theory, observables are defined as (self-adjoint) elements of a
C*-algebra, and states (or, called also expectation values) as normalized
positive linear functionals on the algebra of observables. A simple example of
this sort is given by the well-known Gel'fand isomorphism between a
commutative C*-algebra and a Hausdorff space as its spectrum. In more detail,
denoting the categories of commutative C*-algebras and of Hausdorff spaces,
respectively, by $CommC^{\ast}\!Alg$\ and $HausSp$, we have the following
isomorphic relations between the relevant morphisms in the two categories for
$\mathfrak{A\in}CommC^{\ast}\!Alg$, $M\in HausSp$,
\begin{equation}
CommC^{\ast}\!Alg(\mathfrak{A},C_{0}(M))\simeq HausSp(M,Spec(\mathfrak{A})),
\end{equation}
where $C_{0}(M)$ is the commutative C*-algebra consisting of functions on $M$
vanishing at infinity, and $Spec(\mathfrak{A}):=\{\chi:\mathfrak{A}%
\rightarrow\mathbb{C}$ $|$ $\chi$: $\mathrm{character}\;\mathrm{satisfying}%
\;\chi(AB)=\chi(A)\chi(B)$ for $A,B\in\mathfrak{A}\}$. The isomorphism
$\simeq$ is determined by the equality $[\varphi^{\ast}(x)](A)=[\varphi
(A)](x)$ for a *-homomorphism $\varphi:\mathfrak{A}\rightarrow C_{0}(M)$ and
its dual map $\varphi^{\ast}:M\rightarrow$ $Spec(\mathfrak{A})$. When $M=$
$Spec(\mathfrak{A})$, this reduces to the identification, $\mathfrak{A}\simeq
C_{0}(Spec(\mathfrak{A}))$, between an abstract commutative C*-algebra
$\mathfrak{A}$ and a concrete commutative C*-algebra $C_{0}(Spec(\mathfrak{A}%
))$ of continuous functions on $Spec(\mathfrak{A})$ through the relation
$\mathfrak{A}\ni A\longleftrightarrow\hat{A}\in C_{0}(Spec(\mathfrak{A}))$
defined by $\chi(A)=\hat{A}(\chi)$, $\chi\in Spec(\mathfrak{A})$. In this
connection, a state as an expectation value can be shown just to correspond to
a probability measure according to Markov-Kakutani theorem \cite{BR1}, which
involves already such a statistical aspect as i.i.d. property as will be shown
later. Other examples are given as follows:

\begin{exam}
A finite-dimensional vector space $V$ is isomorphic with its second dual
$V^{\ast\ast}$:
\begin{equation}
V\cong V^{\ast\ast}.
\end{equation}

\end{exam}

\begin{exam}
Let $G$ be a locally compact abelian group. Its dual group $\hat{G}$ defined
by the set of unitary characters on $G$ is also a locally compact abelian
group w.r.t. the pointwise product. Furthermore, $\widehat{(\widehat{G}%
)}=:\widehat{\widehat{G}}$, called the second dual group, can be defined and
the following relation, called Pontryagin duality, holds:
\begin{equation}
\widehat{\widehat{G}}\cong G,
\end{equation}
as topological groups.
\end{exam}

In statistics also, duality is known to play essential roles: the Riemannian
geometric formulation of statistics is called information geometry \cite{AN00}
and based on a duality structure. For any $\alpha\in\mathbb{R}$, the $\alpha
$-connection is defined on the Riemannian manifold consisting of a family of
probability distributions and has the dual connection corresponding to the
$(-\alpha)$-connection. The $\alpha$-connection determines the unique
quasi-distance of probability distributions, called an $\alpha$-divergence,
which generalizes the Kullback-Leibler divergence. Thus the duality is seen to
play essential roles in various contexts.

To consolidate the natural inter-relations between experiments and theories,
and between \textquotedblleft Macro\textquotedblright\ and \textquotedblleft
Micro\textquotedblright, we proceed further to a theoretical framework based
on the Micro-Macro duality. In terms of the four basic ingredients closely
related to the representation-theoretical context of dynamical systems, a
coherent scheme for theoretical description of a target system of our
recognition can be formulated as a pair of duality pairs, which we call a
quadrality scheme \cite{Oj10}:

\vskip10pt

$%
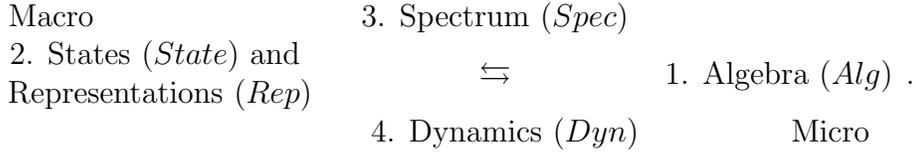
\begin{array}
[c]{ccc}%
\text{Macro \ \ \ \ \ \ \ \ \ \ \ \ \ \ \ \ \ \ \ \ } & \text{3. Spectrum
(}Spec\text{)\newline} & \\%
\begin{array}
[c]{c}%
\text{2. States (}State\text{) and }\\
\text{Representations (}Rep\text{)\newline}%
\end{array}
& \leftrightarrows & \text{1. Algebra (}Alg\text{)\newline}\\
& \text{ }\;\text{4. Dynamics (}Dyn\text{)\newline} & \text{
\ \ \ \ \ \ \ \ \ Micro}%
\end{array}
$. $\;$ \vskip10pt $\;$\newline Here, the dynamics (Dyn) at the bottom creates
an algebra (Alg) of observables to characterize an object system, and the
configurations or structures of the objects in (Alg) are described
mathematically in terms the notion of states (as interfaces between Micro and
Macro)\ and the associated (GNS-)representations of (Alg) which we denote by
(Rep). Roughly speaking, \textquotedblleft Micro\textquotedblright%
\ corresponds to a dynamical system consisting of (Dyn) and (Alg), and
\textquotedblleft Macro\textquotedblright\ to a (co)dynamical one of (Rep) and
(Spec). In the direction from \textquotedblleft Macro\textquotedblright\ to
\textquotedblleft Micro\textquotedblright, we find two arrows, one from the
experimental side to the theoretical one in the form of induction processes,
and another in the operational contexts of controls over the system under
consideration, which should include the aspect of the state preparation
indispensable in conducting experiments. The former one, induction, is
materialized usually on the basis of statistical inference, in combination
with suitable choices of classification schemes. The aspects of control theory
and state preparation are strongly interrelated.

To materialize an induction scheme, we should combine the large deviation
principle (LDP for short) as the mathematical core of statistical inference
with the above quadrality scheme in view of its essential roles in
implementing \textquotedblleft Micro-Macro duality\textquotedblright%
\ indispensable for overcoming the Duhem-Quine no-go theorem. From this
viewpoint, we propose in Section 2 Large Deviation Strategy as a systematic
method of induction, where the importance of statistical inference is
emphasized. Here statistical estimation is no more than the method to analyze
several ingredients such as means, probability distributions and coefficients
of stochastic differential equations, and is fundamentally based on LDP
extended by the quadrality scheme. Stein's lemma and Chernoff bounds in
hypothesis testing are the typical examples in this context. All these
discussions explain the reason why we adopt such naming as Large Deviation
Strategy. After briefly mentioning in Section 3 two example cases of the
application of LDS, we clarify in Section 4 the meaningful and precise
relations between quantum and classical levels in the context of estimation
theory, especially concerning the problem of model selection. In Appendix, the
operational meaning of Tomita's theorem of barycentric decomposition crucial
for the second level of LDS is explained from the viewpoint of a measurement
process. In this way, the theoretical bases of LDS can be found in Micro-Macro
duality \cite{Oj05}, quadrality scheme and LDP extended by the quadrality
scheme. Before going into LDS, it will be instructive to explain the mutual
relations among the relevant tools:

\subsection{Interdependence among statistical inference, Micro-Macro duality,
quadrality scheme and LDS}

The logical relations among the three items including LDS itself is actually a
kind of mutual interdependence in the following sense:

i) statistical inference $\Longrightarrow$ Micro-Macro duality: as Micro-Macro
duality is based on the duality between the inductive and deductive arguments,
it is not possible without the reliable methods and schemes for statistical inference.

ii) Micro-Macro duality $\Longrightarrow$ quadrality scheme: The duality
between (Alg) and (Rep) guarantees the matching between what is to be
described in (Alg) and what to describe by (Rep), which is just the most
important step to resolve the difficulties caused by the no-go theorem of
Duhem-Quine thesis mentioned at the beginning. To attain a meaningful
\textit{interpretation} from the items obtained above, we need to apply some
\textit{classification} to the states and representations in (Rep) according
to some relevant viewpoints, as a result of which we can attain the level of
(Spec) containing all the \textit{classifying parameters} to specify each
configuration realized in (Rep). Then the validity of duality between (Spec)
and (Dyn) allows a universal parametrization of the dynamics, (Dyn), of the object
system in terms of the parameters belonging to (Spec), whose special
case can be found in the familiar parametrization of dynamical map
$t\longmapsto$ $\alpha_{t}$ in terms of a time parameter $t\in\mathbb{R}$.

iii) quadrality scheme $\Longrightarrow$ (an extended form of) LDP: the
standard application of LDP starts from the calculation of a rate function to
measure deviations of empirical data of an observable from its
\textquotedblleft true\textquotedblright\ value (of its average), which is
sometimes called the LDP at the first level \cite{E85}. In view of ii) above,
this corresponds to discussing (Alg) (or, more precisely, a subalgebra of
(Alg) generated by the specific observable under consideration). For the
purpose of statistical inference, however, what is most relevant is that of
such a state as generating a certain definite pattern of empirical data, like
the case of a quantum state yielding a statistical ensemble allowing the Born
formula. This requires us to proceed from (Alg) to (Rep) in the quadrality
scheme in ii) in the context of LDP, which constitutes the main contents of
Sec. 2.3. We try further to extend the scheme to incorporate the level of
(Spec) which enables us to deform and adjust the choice of model spaces in an
optimal way and which we call the LDP third level. Once this is achieved, we
can further proceed to the inference of the dynamical law (Dyn) of the system
under consideration, taking advantage of the duality between (Spec) and (Dyn),
which can be called the LDP fourth level.

iv) extended LDP $\Longrightarrow$ LDS: LDS can be obtained by applying the
above extended scheme of LDP to the context of statistical inference, by means
of which we can attain a full-fledged form of the latter, and hence, we can
re-start i). This loop structure can be easily organized into a helical or
spiral form to deepen the levels of our theoretical descriptions.

\section{Large Deviation Strategy}

\subsection{What is Large Deviation Strategy?}

Now, our Large Deviation Strategy (LDS) is a method of statistical inference
by step-by-step inductions according to the basic idea constituting the large
deviation principle (LDP). We suppose that LDS consists of the following four
levels just in parallel with LDP in its extended form:\newline

\textbf{1st level : Abelian von Neumann algebras} \vspace{-2mm}

\begin{center}
Gel'fand representation, Strong law of large numbers(SLLN)\\[0pt]and
statistical inference on abelian von Neumann algebras
\end{center}

\textbf{2nd level : \textit{States} and \textit{Reps}} \vspace{-2mm}

\begin{center}
Measure-theoretical analysis for noncommutative algebras
\end{center}

\textbf{3rd level : \textit{Spec} and \textit{Alg}} \vspace{-2mm}

\begin{center}
Emergence of space-time from composite systems \\
of internal and external degrees of freedom
\end{center}

\textbf{4th level : \textit{Dyn}}$\;$ From emergence to space-time patterns
and time-series analysis\newline\newline The aim of the first level is to
estimate the spectra of observables and their probability distributions. If we
restrict our attention to mutually consistent observables, this is equivalent
to considering the problem to estimate a spectrum of \textit{abelian von
Neumann subalgebra} generated by the mutually commuting observables. The
obtained information at this stage should help us to restrict the class of
states and representations relevant to the second level, the latter of which
aims at the estimate of states and the associated representations defined on
the algebra of all observables. To proceed to the third level, we consider a
composite system consisting of the object system to be described and of the
macroscopic degrees of freedom arising from the processes of emergence from
the microscopic ones. At the fourth level, we consider the estimate of the
dynamics of the system which will allow us to proceed to the stage of
controlling the object system. The following methods will play central roles
in LDS:\newline

\textbf{I. Large deviation principle} \cite{DZ02,E85}: \vspace{-2mm}

\begin{center}
From probablistic fluctuations to statistical inference
\end{center}

\textbf{II. Tomita decomposition theorem and central decomposition}:
\vspace{-2mm}

\begin{center}
To formulate and use state-valued random variables
\end{center}

\textbf{III. The dual }$\widehat{G}$ \textbf{of a group }$G$ \textbf{and its
crossed products}: \vspace{-2mm}

\begin{center}
To reconstruct Micro from the data of Macro
\end{center}

\textbf{\ IV. Emergence: Condensation associated with spontaneous symmetry }
\vspace{-2mm}

\begin{flushright}
\textbf{breaking (SSB) and phase separation}$\;$ in the direction from Micro
to Macro
\end{flushright}

LDP works effectively at each level of our strategy and provides us with the
information of rate functions in such forms as free energy and relative
entropy, for instance. This information clarifies to which extent a given
quantity in question can deviate from its fiducial point which is called the
\textquotedblleft true\textquotedblright\ value. In this way, LDP is seen to
be essential for statistical inference. As discussed in Sec.2.3, the notion of
state-valued random variables can succesfully be formulated in the use of
Tomita decomposition theorem and central decompositions. In the second level
where states and representations are estimated, this formulation enables us to
analyze them in terms of \textquotedblleft numerical\textquotedblright\ data.
We can also see the necessity of the items in the above III and the processes
of emergence in the third level (in reference to \cite{Oj102} and to the
discussion in the previous section).

\subsection{1st Level: Observables and Abelian Subalgebra}

As stated in the previous subsection, we discuss here the mean and the
probability distribution of an observable. Let $\mathfrak{A}$ be a
C$^{\text{*}}$-algebra, $\psi$ be a state (defined as a normalized positive
linear functional) on $\mathfrak{A}$ and $A$ be an observable to be measured
which is identified by an element of $\mathfrak{A}$. $\mathcal{A}$ denotes the
abelian subalgebra of $\mathfrak{A}$ generated by $A$, and states on
$\mathfrak{A}$ is naturally restricted to $\mathcal{A}$. Therefore, we try to
estimate the appropriate pair $(A,\psi|_{\mathcal{A}})$. The candidate of $A $
comes from the following theorem.

\begin{theo}
An abelian von Neumann algebra $\mathfrak{M}$ on a separable Hilbert space
$\mathfrak{H}$ is generated by one element $X$ (belonging to $\mathfrak{M}$).
\end{theo}

If $X$ is selfadjoint, then we put $A=X$.

For an abelian von Neumann algebra $\mathcal{A}$ and $\psi$ a normal state on
$\mathcal{A}$, the following relations hold:
\begin{align*}
\langle\Omega_{\psi},\pi_{\psi}(A)\Omega_{\psi}\rangle=\; &  \psi(A)=\int
\hat{A}(k)d\nu_{\psi}(k),\\
\pi_{\psi}(\mathcal{A})\ni\pi_{\psi}(A) &  \longleftrightarrow\hat{A}\in
L^{\infty}(K,\nu_{\psi}),\\
\mathfrak{H}_{\psi} &  \cong L^{2}(K,\nu_{\psi})\;\\
\;\hspace{-5mm}(\mathfrak{H}_{\psi}\ni\Omega_{\psi}\! &\longleftrightarrow\;1\in L^{2}(K,\nu_{\psi})),\\
\mathcal{A}_{\ast} &  \cong L^{1}(K,\nu_{\psi}),\!\!\!\!\!\!\!\!\!\;
\end{align*}
where $K$ is a compact Hausdorff space and $\nu_{\psi}$ is a Borel measure on
$K$. Every self-adjoint element $\pi_{\psi}(A)$ of $\pi_{\psi}(\mathcal{A})$
is treated as measure-theoretical $\mathbb{R}$-valued random variable $\hat
{A}$. Thus, we can discuss spectra of observables in the commutative case.

For any $\bar{k}=(k_{1},k_{2},\cdots)\in K^{\mathbb{N}}$ and $A=A^{\ast}%
\in\mathcal{A}$, we define $X_{j}(\bar{k})=k_{j}$ and $\hat{A}_{j}(\bar
{k}):=\hat{A}(X_{j}(\bar{k}))$, we see the validity of

\begin{maco}
$\{\hat{A}_{j}\}$ are independent identically distributed (\textquotedblleft
i.i.d.\textquotedblright) random variables.
\end{maco}

For any measure $m$, let $P_{m}:=$ $m^{\mathbb{N}}$ denote the product measure
of $m$ defined by a countably many tensor power. The following theorem is
known to hold:

\begin{theo}[Cram\'{e}r's theorem \cite{DZ02}] 
Let $\displaystyle{M_{n}(\bar{k}):=\frac
{1}{n}(\hat{A}_{1}(\bar{k})+\cdots+\hat{A}_{n}(\bar{k}))}$ and 
$Q_{n}^{(1)}(\Gamma)$ 
$:=P_{\nu_{\psi}}(M_{n}\in\Gamma)$. Then, $Q_{n}^{(1)}$
satisfies LDP with the rate function $\displaystyle{I_{\psi}(a)=
\sup_{t\in\mathbb{R}}\{at-c_{\psi}(t)\}}$ $\;\displaystyle{\left(  c_{\psi} (t)
=\log\int_{\mathbb{R}}e^{tx}\nu_{\psi}(\hat{A}\in dx)\right)  }$:
\begin{align}
-\inf_{a\in\Gamma^{o}}I_{\psi}(a)  &  \leq\liminf_{n\rightarrow\infty}\frac
{1}{n}\log Q_{n}^{(1)}(\Gamma)\nonumber\\
&  \leq\limsup_{n\rightarrow\infty}\frac{1}{n}\log Q_{n}^{(1)}(\Gamma
)\leq-\inf_{a\in\overline{\Gamma}}I_{\psi}(a)
\end{align}

\end{theo}

By this theorem, we can discuss the convergence rate of the arithmetic means
of observables and estimate \textquotedblleft true\textquotedblright\ means.

As the next step, we give a satisfactory formulation for estimating
probability distributions.

\begin{defi}
A family of probability distributions $\{p(x|w)|w\in W\}$ (with a
$\mathrm{compact}$ set $W\subset\mathbb{R}^{l}$) is called a (statistical)
model if it satisfies the condition that the set\newline$\overline{\{
x\in\mathbb{R}^d | p(x|w)>0\} }$ is independent of $w\in W$.
\end{defi}

\begin{defi}
The probability distribution $p_{\pi,\beta}(x|x^{n})$ defined below is called
a Bayesian escort predictive distribution:
\begin{equation}
p_{\pi,\beta}(x|x^{n})=\langle p(x|w)\rangle_{\pi,\beta}^{x^{n}}
=\frac{\displaystyle{\int p(x|w)\prod_{j=1}^{n}p(x_{j}|w)^{\beta}\pi(w)dw}%
}{\displaystyle{\int\prod_{j=1}^{n}p(x_{j}|w)^{\beta}\pi(w)dw}}, 
\end{equation}
where $\pi(w)$ is a probability distribution (p.d., for short) on $W$ and 
$\beta >0$. 
\end{defi}

We denote by $M_{1}(\Sigma)$ the space of Borel probability measures on a Polish space $\Sigma$.
We define the relative entropy of the probability measure $\nu\in M_1(\Sigma)$
with respect to $\mu\in M_1(\Sigma)$ as
\begin{equation}
D(\nu\Vert\mu)=\left\{
\begin{array}
[c]{c}%
\displaystyle{\int d\nu(\rho)\log\dfrac{d\nu}{d\mu}(\rho)}\;\;\;\;(\nu\ll
\mu)\\
+\infty\;\;\;\;(\mathrm{otherwise}).
\end{array}
\right.
\end{equation}
If there exists a probability measure $\sigma\in M_1(\Sigma)$ such that $\nu,\mu \ll \sigma$,
$D(\nu\Vert\mu)$ is also denoted by $D(q\Vert p)$ where $q:=\dfrac{d\nu}{d\sigma}$ and $p:=\dfrac{d\mu}{d\sigma}$.

\begin{theo}

For $r(\cdot\vert{x^{n}})$as a p.d.-valued function $x^{n}=\{x_{1},\cdots
,x_{n}\}\mapsto r(\cdot\vert{x^{n}})$, its risk function
$\mathcal{R}^{n}(p\Vert r)$ defined by
\begin{equation}
\mathcal{R}^{n}(p\Vert r)=\int\!\!\!\int D(p(\cdot|w)\Vert r(\cdot|x^{n}))
\prod_{j=1}^{n}p(x_{j}|w)^{\beta}dx_{j}\pi(w)dw
\end{equation}
is minimized by the Bayesian escort predictive distribution $p_{\pi,\beta}(x|x^{n})$.
\end{theo}

\begin{proof}
See \cite{Ai75,W1}.
\end{proof}

While there are some more items to be treated in statiscal inference, those
appearing in the next subsection are essentially all what we need in the
second level.

\subsection{2nd Level: States and Representations}

In order to proceed to the second level where states of the algebra of
observables are the target to be evaluated, we need to prepare certain
advanced operator-algebraic setting which is provided by Tomita's theorem of
integral decomposition of states. For the purpose, we first review the notion
of sectors. For a C*-algebra $\mathfrak{A}$ let $E_{\mathfrak{A}}$ be the set
of its states defined by normalized positive linear fuctionals on
$\mathfrak{A}$. A state $\omega\in E_{\mathfrak{A}}$ is called a factor state
if the von Neumann algebra $\pi_{\omega}(\mathfrak{A})^{\prime\prime}$
corresponding to the GNS representation $\{\mathfrak{H}_{\omega},\pi_{\omega
}\}$ is a factor with a trivial center: $\mathfrak{Z}_{\omega}(\mathfrak{A}%
):=\pi_{\omega}(\mathfrak{A})^{\prime\prime}\cap\pi_{\omega}(\mathfrak{A}%
)^{\prime}=\mathbb{C}1_{\mathfrak{H}_{\omega}}$. We denote by $F_{\mathfrak{A}%
}$ the set of all factor states of $\mathfrak{A}$. If $\pi$ is a
representation of $\mathfrak{A}$, then a state $\omega$ of $\mathfrak{A}$ is
said to be $\pi$-normal if there exists a normal state $\rho$ of
$\pi(\mathfrak{A})^{\prime\prime}$ such that
\begin{equation}
\omega(A)=\rho(\pi(A))
\end{equation}
for all $A\in\mathfrak{A}$. Two representations $\pi_{1}$ and $\pi_{2}$ of a
C*-algebra $\mathfrak{A}$ are said to be quasi-equivalent and written as
$\pi_{1}\approx\pi_{2}$, if each $\pi_{1}$-normal state is $\pi_{2}$-normal
and vice versa.

\begin{defi}
[\cite{Oj03}]A sector of C$^{\ast}$-algebra $\mathfrak{A}$ is defined by a
quasi-equivalence class of factor states of $\mathfrak{A}$.
\end{defi}

If $\{\pi,\mathfrak{H}\}$ is a representation of a C$^{\ast}$-algebra
$\mathfrak{A}$, and $n$ is a cardinal, let $n\pi$ denote the representation of
$\mathfrak{A}$ on $\mathfrak{H}^{\oplus n}=\bigoplus_{k=1}^{n}\mathfrak{H}$
defined by
\begin{equation}
n\pi(A)\left(  \bigoplus_{k=1}^{n}\xi_{k}\right)  =\bigoplus_{k=1}^{n}\left(
\pi(A)\xi_{k}\right)  .
\end{equation}
By the following standard theorem in the representation theory of operator
algebras, quasi-equivalence between two representations $\pi_{1}$ and $\pi
_{2}$ can be seen as the isomorphism between the corresponding von Neumann
algebras, $\pi_{1}(\mathfrak{A})^{\prime\prime}$ and $\pi_{2}(\mathfrak{A}%
)^{\prime\prime}$, or as the unitary equivalence of $\pi_{1}$ and $\pi_{2}$ up
to multiplicity:

\begin{theo}[see \cite{BR1}] 
Let $\mathfrak{A}$ be a C$^{\ast}$-algebra and let $\{\pi
_{1},\mathfrak{H}_{1}\}$ and $\{\pi_{2},\mathfrak{H}_{2}\}$ be nondegenerate
representations of $\mathfrak{A}$. The following are equivalent:\newline(1)
$\pi_{1}\approx\pi_{2}$;\newline(2) there exists an isomorphism $\tau:\pi
_{1}(\mathfrak{A})^{\prime\prime}\mapsto\pi_{2}(\mathfrak{A})^{\prime\prime}$
such that $\tau(\pi_{1}(A))=\pi_{2}(A)$ for all $A\in\mathfrak{A}$;\newline(3)
there exist cardinals $n$, $m$, projections $E\in n\pi_{1}(\mathfrak{A}%
)^{\prime}$, $F\in n\pi_{2}(\mathfrak{A})^{\prime}$ and unitary elements
$U:\mathfrak{H}_{1}\mapsto F(\mathfrak{H}_{2}^{\oplus m})$, $V:\mathfrak{H}%
_{2}\mapsto E(\mathfrak{H}_{1}^{\oplus n})$ such that
\begin{align*}
U\pi_{1}(A)U^{\ast}  &  =m\pi_{2}(A)F,\\
V\pi_{2}(A)V^{\ast}  &  =n\pi_{1}(A)E
\end{align*}
for all $A\in\mathfrak{A}$;\newline(4) There exists a cardinal $n$ such that
$n\pi_{1}\cong n\pi_{2}$, i.e., $\pi_{1}$ and $\pi_{2}$ are unitary equivalent
up to multiplicity.
\end{theo}

The Gel'fand spectrum $Spec(\mathfrak{Z}_{\omega}(\mathfrak{A}))$ of the
center $\mathfrak{Z}_{\omega}(\mathfrak{A})$ is then identified with a factor
spectrum $\overset{\smallfrown}{\mathfrak{A}}$ of $\mathfrak{A}$:
\[
Spec(\mathfrak{Z}_{\omega}(\mathfrak{A}))\cong\overset{\smallfrown
}{\mathfrak{A}}:=F_{\mathfrak{A}}/\!\approx\;:\mathrm{factor}%
\;\mathrm{spectrum}.
\]
The center $\mathfrak{Z}_{\omega}(\mathfrak{A})$ and the factor spectrum
$\overset{\smallfrown}{\mathfrak{A}}$ play the role of the abelian algebra of
macroscopic order parameters to specify sectors and the classifying space of
sectors to distinguish among different sectors, respectively \cite{Oj03}.

As already mentioned, we need to treat states as objects to be estimated in
the second level of LDP, which means the necessity for \textquotedblleft
states to be treated as observables\textquotedblright. The notion of
state-valued random variables required for this purpose can succesfully be
formulated in the use of Tomita's theorem for orthogonal decompositions of
states by barycentric orthogonal measures whose special case of central
decompositions \cite{Oj98, Oj03} is seen to be particularly useful for our
purposes of statistical inference of state estimate. Now the orthogonality
$\omega_{1}\perp\omega_{2}$ of positive linear functionals $\omega_{i}%
\in\mathfrak{A}_{+}^{\ast}$ and the orthogonal measures $\mu$ on the state
space $E_{\mathfrak{A}}$ of a C*-algebra $\mathfrak{A}$ are defined,
respectively, as follows (see \cite{BR1}):

\begin{defi}
If $\omega_{1},\omega_{2}\in\mathfrak{A}_{+}^{\ast}$ satisfy any of the
following three equivalent conditions, they are said to be \textit{orthognal}
and we write $\omega_{1}\perp\omega_{2}$:
\end{defi}

\begin{enumerate}
\item if $\omega^{\prime}\leq\omega_{1}$ and $\omega^{\prime}\leq\omega_{2}$
for $\omega^{\prime}\in\mathfrak{A}_{+}^{\ast}$ then $\omega^{\prime}=0$;

\item there is a projection $P\in\pi_{\omega}(\mathfrak{A})^{\prime}$ s.t.
$\omega_{1}(A)=\langle P\Omega_{\omega},\pi_{\omega}(A)\Omega_{\omega}\rangle$
and $\omega_{2}(A)=\langle(1-P)\Omega_{\omega},\pi_{\omega}(A)\Omega_{\omega
}\rangle$;

\item the representation associated to $\omega=\omega_{1}+\omega_{2}$ is a
direct sum of the representations associated with $\omega_{1}$ and $\omega
_{2}$,%
\[
\mathfrak{H}_{\omega}=\mathfrak{H}_{\omega_{1}}\oplus\mathfrak{H}_{\omega_{2}%
},\text{ \ }\pi_{\omega}=\pi_{\omega_{1}}\oplus\pi_{\omega_{2}},\text{
\ }\Omega_{\omega}=\Omega_{\omega_{1}}\oplus\Omega_{\omega_{2}}.
\]

\end{enumerate}

\begin{defi}
A positive regular Borel measure $\mu$ on $E_{\mathfrak{A}}$ is defined to be
an \textit{orthogonal measure} on $E_{\mathfrak{A}}$ if it satisfies for any
Borel set $S\subset E_{\mathfrak{A}}$ the condition
\begin{equation}
\left(  \int_{S}\rho\;d\mu(\rho)\right)  \perp\left(  \int_{E_{\mathfrak{A}%
}\backslash S}\rho\;d\mu(\rho)\right)  .
\end{equation}

\end{defi}

The important properties characteristic to these notions can be found in the
following theorem due to Tomita:

\begin{theo}[Tomita's theorem, see \cite{BR1}] 
Let $\mathfrak{A}$ be a C$^{\ast}$-algebra
 and $\omega$ be a state on $\mathfrak{A}$. There exists one-to-one
correspondence between the following three sets$:$\newline$(1)$ the orthogonal
measures $\mu$ with barycenter $\displaystyle{\omega=\int_{E_{\mathfrak{A}}%
}\rho d\mu(\rho)};$\newline$(2)$ the abelian von Neumann subalgebras
$\mathfrak{B}\subseteq\pi_{\omega}(\mathfrak{A})^{\prime};$\newline$(3)$ the
orthogonal projections $P$ on $\mathfrak{H}_{\omega}$ such that
\[
P\Omega_{\omega}=\Omega_{\omega},\text{ \ \ \ }P\pi_{\omega}(\mathfrak{A}%
)P\subseteq\{P\pi_{\omega}(\mathfrak{A})P\}^{\prime}%
\]
If $\mu,\mathfrak{B},P$ are in correspodence one has the following
relations$:$\newline$(1)$ $\mathfrak{B}=\{\pi_{\omega}(\mathfrak{A})\cup
P\}^{\prime};$ $(2)$ $P=[\mathfrak{B}\Omega_{\omega}];$\newline$(3)$
$\mu(\widehat{A}_{1}\widehat{A}_{2}\cdots\widehat{A}_{n})=\langle
\Omega_{\omega},\pi_{\omega}(A_{1})P\pi_{\omega}(A_{2})P\cdots P\pi_{\omega
}(A_{n})\Omega_{\omega}\rangle;$\newline$(4)$ $\mathfrak{B}$ is $\ast
$-isomorphic to the range of the map $L^{\infty}(\mu):=L^{\infty
}(E_{\mathfrak{A}},\mu) \ni f\mapsto\kappa_{\mu}(f)\in\pi_{\omega
}(\mathfrak{A})^{\prime}$ defined by
\begin{equation}
\langle\Omega_{\omega},\kappa_{\mu}(f)\pi_{\omega}(A)\Omega_{\omega}%
\rangle=\int_{E_{\mathfrak{A}}}f(\rho)\widehat{A}(\rho)d\mu(\rho)
\end{equation}
and for $A,B\in\mathfrak{A}$
\begin{equation}
\kappa_{\mu}(\widehat{A})\pi_{\omega}(B)\Omega_{\omega}=\pi_{\omega}%
(B)P\pi_{\omega}(A)\Omega_{\omega},
\end{equation}
where the map $\mathfrak{A}\ni A\longmapsto$ $\widehat{A}\in L^{\infty}(\mu)$
is defined by $\widehat{A}$ $:=(E_{\mathfrak{A}}\ni\rho\longmapsto$ $\rho(A))$.
\end{theo}

The above measure $\mu$ is called a barycentric measure of the state $\omega$,
which is, in turn, called the barycenter $\displaystyle{\omega=b(\mu
):=\int_{E_{\mathfrak{A}}}\rho\;d\mu(\rho)}$ of $\mu$. The set of orthogonal
probability measures $\mu$ on $E_{\mathfrak{A}}$ with barycentre $\omega$ is
denoted by $\mathcal{O}_{\omega}(E_{\mathfrak{A}})$. In reference to the
abelian von Neumann algebra $\mathfrak{B}$, we also denote the measure $\mu$
by $\mu_{\mathfrak{B}}$. We add here the following observation to extend the
essential contents of the Gel'fand isomorphism for commutative C*-algebras to
the non-commutative situation: the image $\widehat{\mathfrak{A}}%
:=\{\widehat{A}|A\in\mathfrak{A}\}$ of the map $\mathfrak{A}\ni A\longmapsto$
$\widehat{A}\in L^{\infty}(\mu)$ is contained in the universal enveloping von
Neumann algebra $\mathfrak{A}^{\ast\ast}$ of $\mathfrak{A}$ and constitutes a
C*-algebra of measure-theoretical random variables equipped with a linear
structure $(\alpha\widehat{A}+\beta\widehat{B})(\rho)$ $:=(\widehat{\alpha
A+\beta B})(\rho)$ ($\alpha,\beta\in\mathbb{C}$), a non-commutative
convolution product defined by $(\widehat{A}\ast\widehat{B})(\rho)$
$:=\widehat{AB}(\rho)$, and the norm $\Vert\cdot\Vert$ given by
\begin{equation}
\Vert\widehat{A}\Vert=\sup_{%
%TCIMACRO{\QATOP{\rho\in E_{\QTR{frak}{A}},}{\Vert\rho\Vert=1}}%
%BeginExpansion
\genfrac{}{}{0pt}{}{\rho\in E_{\mathfrak{A}},}{\Vert\rho\Vert=1}%
%EndExpansion
}|\widehat{A}(\rho)|.
\end{equation}

\begin{defi}
If the algebra $\mathfrak{B}$ corresponding to $\mu$ is a subalgebra of the
center $\mathfrak{Z}_{\omega}(\mathfrak{A})$ of the GNS representation
$\pi_{\omega}$ of $\omega$, the orthogonal measure $\mu=\mu_{\mathfrak{B}}%
\in\mathcal{O}_{\omega}(E_{\mathfrak{A}})$ is called a subcentral measure of
$\omega$ satisfying the condition that, for any Borel set $\Delta\subset
E_{\mathfrak{A}}$, the pair of subrepresentations, $\displaystyle{\int
_{\Delta}^{\oplus} \pi_{\rho}\;d\mu(\rho)}$ and $\displaystyle{\int
_{E_{\mathfrak{A}}\backslash\Delta}^{\oplus} \pi_{\rho}\;d\mu(\rho)}$, of
$\pi_{\omega}$ are disjoint in the sense of the absence of non-zero
intertwiners. In the case of $\mathfrak{B}=\mathfrak{Z}_{\omega}%
(\mathfrak{A})$, the corresponding subcentral measure is called the central
measure of $\omega$ and denoted by $\mu_{\omega}:=\mu_{\mathfrak{Z}_{\omega
}(\mathfrak{A})}\in\mathcal{O}_{\omega}(E_{\mathfrak{A}})$.
\end{defi}

Since $\kappa_{\mu_{\omega}}$ is a *-algebraic embedding of $L^{\infty}(\mu)$,
we can define a projection-valued measure (PVM) $E_{\omega}:(\mathfrak{B}%
(\mathrm{supp}\;\mu_{\omega})\ni\Delta\mapsto E_{\omega}(\Delta)\in
Proj(\mathfrak{Z}_{\omega}(\mathfrak{A})))$ on Borel subsets $\Delta
\in\mathfrak{B}(\mathrm{supp}\;\mu_{\omega})$ of the state space
$E_{\mathfrak{A}}$ by $E_{\omega}(\Delta):=\kappa_{\mu_{\omega}}(\chi_{\Delta
})\in Proj(\mathfrak{Z}_{\omega}(\mathfrak{A})))$, which satisfies%

\begin{equation}
\label{centralPVM}\langle\Omega_{\omega},E_{\omega}(\Delta)\Omega_{\omega
}\rangle=\mu_{\omega}(\Delta).
\end{equation}
Here the indicator function $\chi_{\Delta}$ for a subset $\Delta$ of
$E_{\mathfrak{A}}$ is defined as usual:
\[
\chi_{\Delta}(\rho)=\left\{
\begin{array}
[c]{c}%
1\;\;\;\;(\rho\in\Delta),\\
0\;\;\;\;(\rho\notin\Delta).
\end{array}
\right.
\]
In this way, states $\rho$ on $\mathrm{supp}(\mu_{\omega})$ constitute a
random variable on the central spectrum, and each element $\kappa_{\mu
_{\omega}}(f)$ $\in\kappa_{\mu_{\omega}}(L^{\infty}(\mu_{\omega}%
))=\mathfrak{B}=\mathfrak{Z}_{\omega}(\mathfrak{A})$ can be expressed as
\begin{equation}
\kappa_{\mu_{\omega}}(f)=\int f(\rho)dE_{\omega}(\rho).
\end{equation}
Therefore, the center $\mathfrak{Z}_{\omega}(\mathfrak{A})$ of $\mathfrak{A} $
can be seen as an algebra consisting of non-linear functions of states.

When the methods discussed in this section are applied to practical
situations, it will be safe and also sufficient for us to restrict ourselves
to such cases that the support of the barycentric measure $\mu_{\omega}$ is a
compact subset $B$ in the factor spectrum $F_{\mathfrak{A}}$ of $\mathfrak{A}%
$:
\begin{equation}
\omega=\int\rho\;d\mu_{\omega}(\rho)=\int_{B}\rho_{\xi}\;d\tilde{\mu}(\xi)
\end{equation}
where $\{\rho_{\xi}|\xi\in\Xi:\mathrm{an}\;\mathrm{order}\;\mathrm{parameter}%
\}\subset F_{\mathfrak{A}}$. Here the factor spectrum $F_{\mathfrak{A}}$ of
$\mathfrak{A}$ means the subset of the state space $E_{\mathfrak{A}}$
consisting of all the factor states $\varphi$ whose GNS representations have
trivial centers: $\mathfrak{Z}_{\varphi}(\mathfrak{A})=\pi_{\varphi
}(\mathfrak{A})^{\prime\prime}\cap\pi_{\varphi}(\mathfrak{A})^{\prime
}=\mathbb{C}1_{\mathfrak{H}_{\varphi}}$.

\subsubsection{Mathematical and statistical basis}

Let $\mathfrak{A}$ be a separable C$^{\ast}$-algebra and $\psi$ be a state
on $\mathfrak{A}$. Then $E_{\mathfrak{A}}$ is weak $\ast$-compact and
metrizable by the metric
\begin{equation}
\label{distance}d(\omega_{1},\omega_{2})=\sum_{j=1}^{\infty}\frac{1}{2^{j}%
}\frac{\left\vert \omega_{1}(A_{j})-\omega_{2}(A_{j})\right\vert }{\Vert
A_{j}\Vert} ,
\end{equation}
where the set $\{A_{j}\in\mathfrak{A}\vert A_{j}\neq0, j=1,2,\cdots\}$ is a
dense subset of $E_{\mathfrak{A} }$. Thus $\mathrm{supp}\;\mu_{\psi}$ of the
central measure $\mu_{\psi}$ of $\psi\in E_{\mathfrak{A}}$ is compact in the
weak $\ast$-topology and $(\mathrm{supp}\;\mu_{\psi})^{\mathbb{N}}$ is also
compact by Tikhonov's theorem. For $\tilde{\rho}=(\rho_{1},\rho_{2},\cdots
)\in(\mathrm{supp}\;\mu_{\psi})^{\mathbb{N}}$, we define $Y_{j}(\tilde{\rho
})=\rho_{j}$. Each $\rho_{j}$ is a factor state because $\mu_{\psi}$ is
supported by the closed subset of $F_{\mathfrak{A}}$, the set of factor
states. Then $\{Y_{j}\}_{j=1}^{\infty}$ is seen to be a set of ($\mathrm{supp}%
\;\mu_{\psi}$)-valued random variables satisfying the following condition:

\begin{maco}
$\{Y_{j}\}$ are independent identically distributed (\textquotedblleft
i.i.d.\textquotedblright) random variables.
\end{maco}

We denote by $M_{1}(\Sigma)$ the space of Borel probability measures on a Polish space $\Sigma$,
and by $B(\Sigma)$ the vector space of all bounded
Borel measurable functions on $\Sigma$, respectively. For $\phi\in B(\Sigma)$,
let $\tau_{\phi}:M_{1}(\Sigma)\rightarrow\mathbb{R}$ be defined by $\tau
_{\phi}(\nu)=\langle\phi,\nu\rangle=\displaystyle{\int_{\Sigma}\phi\;d\nu}$.
We denote by $\mathcal{B}_{cy}(M_{1}(\Sigma))$ the $\sigma$-field of cylinder
sets on $M_{1}(\Sigma)$, i.e., the smallest $\sigma$-field that makes all
$\{\tau_{\phi}\}$ measurable (see \cite{DZ02}).

For any $\tilde{\rho}\in(\mathrm{supp}\;\mu_{\psi})^{\mathbb{N}}$,
$A\in\mathcal{B}(\mathrm{supp}\;\mu_{\psi})$ and $\Gamma\in\mathcal{B}%
_{cy}(M_{1}(E_{\mathfrak{A}}))$, we define the empirical measures
\begin{equation}
L_{n}(\tilde{\rho},A)=\frac{1}{n}\sum_{j=1}^{n}\delta_{Y_{j}(\tilde{\rho}%
)}(A),
\end{equation}
and
\begin{equation}
Q_{n}^{(2)}(\Gamma)=P_{\mu_{\psi}}(L_{n}\in\Gamma).
\end{equation}

The next theorem \cite{HOT83} is the key to proving LDP:

\begin{theo}[HOT83] \label{HOT83}
Let $\mu,\nu$ be regular Borel probability measures on $E_{\mathfrak{A}%
}$ with\newline barycenters $\omega,\psi$ $\in E_{\mathfrak{A}}$. If there is
a subcentral measure $m$ on $E_{\mathfrak{A}}$ such that $\mu,\nu$ $\ll m$,
then $S(\psi\Vert\omega)$ $=D(\nu\Vert\mu)$.
\end{theo}

This theorem enables us to evaluate the quantum relative entropy
$S(\psi\Vert\omega)$ as the measure-theoretical relative entropy $D(\nu\Vert\mu)$.

\begin{theo}
Let $\mathfrak{A}$ be a separable C$^{\ast}$-algebra and $\psi$ be a state
on $\mathfrak{A}$. Then $Q_{n}^{(2)}$ satisfies LDP with the rate function
$S(b(\cdot)\Vert\psi)$:
\begin{equation}
-\!\!\!\inf_{\nu\in\Gamma^{o},\nu \ll \mu_\psi}\!\!\!
S(b(\nu)\Vert\psi) \leq\liminf_{n\rightarrow\infty}\frac{1}{n}\log Q_{n}^{(2)}(\Gamma)
\leq\limsup_{n\rightarrow\infty}\frac{1}{n}\log Q_{n}^{(2)}(\Gamma)
\leq-\!\!\!\inf_{\nu\in\overline{\Gamma},\nu \ll \mu_\psi}\!\!\!S(b(\nu)\Vert\psi)
\end{equation}
for any $\Gamma\in\mathcal{B}_{cy}(M_{1}(E_{\mathfrak{A}}))$.
In the case that, for $\Gamma\in\mathcal{B}_{cy}(M_{1}(E_{\mathfrak{A}}))$,
$\{\nu\in\Gamma^{o}|\nu \ll \mu_\psi\}$ and $\{\nu\in\overline{\Gamma}|\nu \ll \mu_\psi\}$
are empty, $\displaystyle{\inf_{\nu\in\Gamma^{o},\nu \ll \mu_\psi}\!\!\!S(b(\nu)\Vert\psi)}$ and
$\displaystyle{\inf_{\nu\in\overline{\Gamma},\nu \ll \mu_\psi}\!\!\!S(b(\nu)\Vert\psi)}$ are
defined as infinity, respectively.
\end{theo}

\begin{proof}
$(E_{_{\mathfrak{A}}},d)$ with the metric $d$ defined by (\ref{distance}) is a
compact metric space, so is a Polish one. Therefore, we can apply Sanov's
theorem \cite{DZ02} for $Q_{n}^{(2)}(\Gamma)$ to prove this theorem by using
Theorem \ref{HOT83} (HOT83).
\end{proof}

In Appendix A a generalization of this theorem will be discussed, but we
should first justify the use of generic barycentric measures in the context of
statistical inference.

\begin{defi}
A family of states $\{\omega_{\theta}|\theta\in\Theta\}$ parametrized by a
compact set $\Theta$ in $\mathbb{R}^{d}$ is called a (statistical) model if it
satisfies the following three conditions: \newline$\;\;(i)$ There is a
subcentral measure $m$ on $E_{\mathfrak{A}}$ such that $\mu_{\omega_{\theta}%
}\ll m$ for every $\theta\in\Theta$.\newline$\;(ii)$ The set
$\displaystyle{\overline{\left\{  \rho\in E_{\mathfrak{A}}\Big|p_{\theta}:=\frac
{d\mu_{\omega_{\theta}}}{dm}(\rho)>0\right\}  }}$ is independent of $\theta
\in\Theta$.\newline$(iii)$ $\omega_{\theta}$ is Bochner integrable.
\end{defi}

\begin{defi}
For a given model $\{\omega_{\theta}\}_{\theta\in\Theta}$, a probability
distributions $\pi(\theta)$ of $\theta$, a state
$\omega_{\pi,\beta}^{n}$ defined by
\begin{equation}
\omega_{\pi,\beta}^{n}:=\frac{\displaystyle{\int\omega_{\theta}\prod
_{j=1}^{n}p_{\theta}(\rho_j)^{\beta}\pi(\theta)d\theta}}{\displaystyle{\int
\prod_{j=1}^{n}p_{\theta}(\rho_j)^{\beta}\pi(\theta)d\theta}}%
\end{equation}
is called a Bayesian escort predictive state, where $\rho^{n}:=\{\rho_{1}%
,\cdots,\rho_{n}\}$ and $\beta>0$.
\end{defi}

We obtain the following important theorem:

\begin{theo}
For $\phi^{\rho^{n}}$as a state-valued function $\rho^{n}=\{\rho_{1},\cdots
,\rho_{n}\}\mapsto\phi^{\rho^{n}}\in E_{\mathfrak{A}}$, its risk function
$T^{n}(\omega_{\theta}\Vert\phi^{\rho^{n}})$ defined by%
\begin{align}
T^{n}(\omega_{\theta}\Vert\phi^{\rho^{n}})& :=\frac{1}{A}\int\!\!\!\int
S(\omega_{\theta}\Vert\psi^{\rho^{n}})\prod_{j=1}^{n}p_{\theta}(\rho_{j})^{\beta
}dm(\rho_{j})\pi(\theta)d\theta,\\
A  &  :=\int\!\!\!\int\prod_{j=1}^{n}p_{\theta}(\rho_{j})^{\beta}dm(\rho_{j}%
)\pi(\theta)d\theta,\nonumber
\end{align}
is minimized by the Bayesian escort predictive state $\omega_{\pi,\beta
}^{n}$.
\end{theo}

\begin{proof}
For any measure $\mu=\mu^{\rho^{n}}$ on $E_{\mathfrak{A}}$ with $\phi^{\rho^{n}} $
as its barycenter such that $\mu\ll m$, we have
\begin{align}
S(\omega_{\theta}\Vert\psi^{\rho^{n}}) & =D(\mu_{\omega_{\theta}}\Vert\mu)\nonumber\\
&  =\int dm\frac{d\mu_{\omega_{\theta}}}{dm}\left(  \log\frac{d\mu
_{\omega_{\theta}}}{dm}-\log\frac{d\mu}{dm}\right)  ,\nonumber
\end{align}
and hence,
\begin{align}
A(T^{n}(\omega_{\theta}\Vert\phi_1^{\rho^{n}})-T^{n}(\omega_{\theta}\Vert\phi_2^{\rho^{n}})) & \;\nonumber\\
=\int\!\!\!\int\!\!\!\int dm\frac{d\mu_{\omega_{\theta}}}{dm}\left(  \log
\frac{d\mu_{2}^{\rho^{n}}}{dm}-\log\frac{d\mu_{1}^{\rho^{n}}}{dm}\right)   &
\;\prod_{j=1}^{n}p_{\theta}(\rho_{j})^{\beta}dm(\rho_{j})\pi(\theta)d\theta
\nonumber\\
=\int\!\!\!\int dm\left(  \int\frac{d\mu_{\omega_{\theta}}}{dm}\prod_{j=1}^{n}
p_{\theta}(\rho_{j})^{\beta}\pi(\theta)d\theta\right)   &  \;\left(
\log\frac{d\mu_{2}^{\rho^{n}}}{dm}-\log\frac{d\mu_{1}^{\rho^{n}}}{dm}\right)
\prod_{j=1}^{n}dm(\rho_{j}),\nonumber
\end{align}
for any $\phi_{1}^{\rho^{n}}=b(\mu_{1}^{\rho^{n}}),\phi_{2}^{\rho^{n}}=b(\mu_{2}%
^{\rho^{n}})\in E_{\mathfrak{A}}$ such that $\mu_{1}^{\rho^{n}},\mu_{2}^{\rho^{n}}\ll
m$. Now we put $\displaystyle{\tau:=\frac{1}{B}\int\mu_{\omega_{\theta}}}$
$\displaystyle{\prod_{j=1}^{n}p_{\theta}(\rho_{j})^{\beta}\pi(\theta)d\theta}$
with $\displaystyle{B:=\int\prod_{j=1}^{n} p_{\theta}(\rho_{j})^{\beta}\pi
(\theta)d\theta}$. If $\dfrac{d\mu_{2}^{\rho^{n}}}{dm}$ is equal to $\dfrac
{d\tau}{dm}$, then the above equalities continue to the following form:
\begin{align}
&  =B\int\!\!\!\int D(\tau\Vert\mu_{1}^{\rho^{n}})\prod_{j=1}^{n}dm(\rho_{j})\nonumber\\
&  =B\int\!\!\!\int S(b(\tau)\Vert\phi_{1}^{\rho^{n}})\prod_{j=1}^{n}dm(\rho_{j})\geq 0\nonumber
\end{align}
for any $\phi_{1}^{\rho^{n}}\in E_{\mathfrak{A}}$. Therefore, the risk function
$T^{n}(\omega_{\theta}\Vert\phi^{\rho^{n}})$ is minimized at the unique state
$\phi^{\rho^{n}}=b(\tau)=\omega_{\pi,\beta}^{n}$.
\end{proof}

This theorem is a generalization of \cite{Ai75} and \cite{TK05}, and explains
the reason why the Bayesian escort predictive state is a good estimator for a
\textquotedblleft true\textquotedblright$\;$one.

Now we discuss the situations with singular statistics. The results here are
proved originally in \cite{W1,W2,W3}. The reason why we use this method will
be explained in Section 4. For a model $\{\omega_{\theta}\}_{\theta\in\Theta}$
and a \textquotedblleft true\textquotedblright$\;$state $\psi\in
E_{\mathfrak{A}}$ we assume that there is a subcentral measure $m$ satisfying
$\mu_{\omega_{\theta}},\mu_{\psi}\ll m$ and
\[
\overline{\left\{  \rho\in E_{\mathfrak{A}}\Big|\;p(\rho|\theta):=p_{\theta
}(\rho)=\dfrac{d\mu_{\omega_{\theta}}}{dm}(\rho)>0\right\}  }=\overline
{\left\{  \rho\in E_{\mathfrak{A} }\Big|\;q(\rho):=\dfrac{d\mu_{\psi}}%
{dm}(\rho)>0\right\}  }%
\]
for every $\theta\in\Theta$, and we consider%

\begin{equation}
L(\theta):=-\int dm(\rho)q(\rho)\log p(\rho|\theta).
\end{equation}

We assume that there exists at least one parameter $\theta\in\Theta$ that
minimizes $L(\theta)$,
\begin{equation}
L_{0}=\min_{\theta\in\Theta}L(\theta)
\end{equation}
and that $p_{0}(\rho):=p(\rho|\theta_{0})$ is one and the same density
function for any $\theta_{0}\in\Theta_{0}:$ $=\{\theta\in\Theta|L(\theta
)=L_{0}\}$, and we put $\omega_{0}:=\omega_{\theta_{0}}$. Then, from such
definitions as
\begin{align}
f(\rho,\theta)  &  :=\log\frac{p_{0}(\rho)}{p(\rho|\theta)},\label{f}\\
D(\theta)  &  :=\int dm(\rho)q(\rho)f(\rho,\theta),\label{D}\\
D_{n}(\theta)  &  :=\frac{1}{n}\sum_{j=1}^{n}f(\rho_{j},\theta), \label{D_n}%
\end{align}
it immediately follows that
\begin{align}
D(\theta)  &  =S(\psi\Vert\omega_{\theta})-S(\psi\Vert\omega_{0}),\\
D_{n}(\theta)  &  =S_{n}(\psi\Vert\omega_{\theta})-S_{n}(\psi\Vert\omega_{0}),
\end{align}
where $\displaystyle{S_{n}(\psi\Vert\omega_{\theta})=\dfrac{1}{n}\sum
_{j=1}^{n}\log\dfrac{q(\rho_{j})}{p(\rho_{j}|\theta)}}$. Therefore, we see
$D(\theta)\geq0$.\newline\textbf{Assumptions.} $(1)$ The open kernel
$\Theta^{o}$ of the set $\Theta$ of parameters $\theta$ is non-empty. The
boundary of $\Theta$ is defined by real analytic functions $\varrho_{j}%
(\theta)$ so that
\begin{equation}
\Theta=\{\theta\in\mathbb{R}^{d}|\varrho_{1}(\theta)\geq0,\varrho_{2}%
(\theta)\geq0,\cdots,\varrho_{k}(\theta)\geq0\}.
\end{equation}

$(2)$ The a priori distribution $\pi(\theta)$ is factorized into the product,
\begin{equation}
\pi(\theta)=\pi_{1}(\theta)\pi_{2}(\theta),
\end{equation}
\ of a real analytic function $\pi_{1}(\theta)\geq0$ and of a function of C$^{\infty}$-class 
\ $\pi_{2}(\theta)>0$.

$(3)$ The map $\Theta\ni\theta\mapsto f(\rho,\theta)$ is an $L^{s}(q)$-valued
analytic function, where $L^{s}(q)$ with $s\geq6$ is defined by
\begin{equation}
L^{s}(q):=\left\{  f(\rho)\left\vert \Vert f\Vert_{s}:=\left(  \int\left\vert
f(\rho)\right\vert ^{s}q(\rho)dm(\rho)\right)  ^{1/s}<\infty\right.  \right\}
.
\end{equation}

$(4)$ There is an $\epsilon>0$ such that
\begin{equation}
\int\left(  \sup_{\theta\in\Theta}\left\vert f(\rho,\theta)\right\vert
^{2}\right)  \left(  \sup_{D(\theta)<\epsilon}p(\rho|\theta)\right)
dm(\rho)<\infty.
\end{equation}
\vspace{2mm}

The pair $(\psi,\omega_{\theta})$ is said to be coherent if there exist $A>0$
and $\epsilon>0$ such that
\begin{equation}
\theta\in\Theta_{\epsilon}\Rightarrow S(\psi\Vert\omega_{\theta})-S(\psi\Vert\omega_{0})
\geq A\cdot S(\omega_{0}\Vert\omega_{\theta}),
\end{equation}
where $\Theta_{\epsilon}:=\left\{  \theta\in\Theta\left\vert S(\omega_{0}\Vert\omega_{\theta})
\leq\epsilon\right.  \right\}  $ and, otherwise, incoherent.

In the rest of this subsection, we assume the following condition to hold, in
addition to the above assumptions (1) -- (4):

\begin{maco}
The pair $(\psi,\omega_{\theta})$ satisfies the coherence condition.
\end{maco}

We note that the validity of these assumptions means the interplay between
positivity and analyticity which is closely related with the modular structure
inherent in the standard form of a von Neumann algebra.

The inequality in (34) can be written as follows:
\begin{equation}
\int dm(\rho)q(\rho)\log\frac{p_{0}(\rho)}{p(\rho|\theta)}\geq A\cdot
D(p_{0}\Vert p_{\theta})
\end{equation}
for every $\theta\in\Theta_{\epsilon}$, and the following inequality holds:
\begin{equation}
t+e^{-t}-1\geq B(\eta)t^{2}%
\end{equation}
for $|t|<\eta$, where $B(\eta)$ is a monotone decreasing strictly positive
function of $\eta>0$. Thus, by fixing $\eta$ sufficiently large, it holds
that
\begin{equation}
L(\theta)-L_{0}\geq C\int dm(\rho)p_{0}(\rho)f(\rho,\theta)^{2}\quad(C>0),
\end{equation}
for every $\theta\in\Theta_{\epsilon}$. We can prove the next theorem by using
the same methods as in \cite{W3}:

\begin{theo}
\label{standardform} By the resolution of singularities, the functions in Eqs.
(\ref{D}), (\ref{f}), (\ref{D_n}) can be reduced to the following
\textquotedblleft standard forms\textquotedblright:
\begin{align}
D(g(u))  &  =u^{2k}=u_{1}^{2k_{1}}\dots u_{d}^{2k_{d}},\\
f(\rho,g(u))  &  =a(\rho,u)u^{k},\\
D_{n}(g(u))  &  =u^{2k}-\frac{1}{\sqrt{n}}u^{k}\xi_{n}(u),
\end{align}
where $u=(u_{1},\cdots,u_{d})$ is a coordinate system of an analytic manifold
$U$, and $g$ is an analytic map from $U$ to $\Theta$, $k_{1},\cdots,k_{d}$ are
non-negative integers, $a(\rho,u)$ is an analytic function on $U$ for each
$\rho\in\mathrm{supp}\;\mu_{\omega_{\theta}}$ such that $E_{\rho}%
[a(\rho,u)]=u^{k}$, and $\{\xi_{n}\}$ is an empirical process such that
\begin{equation}
\xi_{n}(u)=\frac{1}{\sqrt{n}}\sum_{j=1}^{n}\{a(\rho_{j},u)-u^{k}\},
\end{equation}
converges weakly to the Gaussian process $\xi(u)$ with expectation $E_{\xi}[\xi(u)]=0$ and
covariance $E_{\xi}[\xi(u)\xi(v)]=E_\rho[a(\rho,u)a(\rho,v)]-u^kv^k$.
\end{theo}

The universal validity of the above standard forms of the quantum relative
entropy and of the log likelihood ratio function, respectively, is guranteed
independently of the model by this theorem.

Furthermore, we need the following theorem proved in \cite{W1}, as a next step 
to Theorem \ref{standardform}, under the above assumptions:

\begin{theo}[\cite{W1}]
 $(1)$ The set of parameters $\mathcal{U}=g^{-1}(\Theta_{\epsilon
})$ is covered by a finite set
\[
\mathcal{U}=\bigcup_{\alpha}U_{\alpha},
\]
where $U_{\alpha}$ is given by a local coordinate,
\[
U_{\alpha}=[0,b]^{d}=\{(u_{1},u_{2},\cdots,u_{d})\;|\;0\leq u_{1},u_{2}%
,\cdots,u_{d}\leq b\}.
\]
$(2)$ In each $U_{\alpha}$,
\[
D(g(u))=u^{2k}=u_{1}^{2k_{1}}\dots u_{d}^{2k_{d}},
\]
where $k_{1},\cdots,k_{d}$ are non-negative integers.\newline$(3)$ There is a
positive function $\tilde{\pi}(u)$ of class C$^{\infty}$ such that
\begin{equation}
\pi(g(u))|g^{\prime}(u)|=\tilde{\pi}(u)u^{h}=\tilde{\pi}(u)u_{1}^{h_{1}}%
u_{2}^{h_{2}}\cdots u_{d}^{h_{d}},
\end{equation}
where $|g^{\prime}(u)|$ is the absolute value of the Jacobian determinant and
$h_{1},\cdots,h_{d}$ are non-negative integers and
\[
\tilde{\pi}(u)>c>0,
\]
is a function of class C$^{\infty}$, where $c>0$ is a positive
constant.\newline$(4)$ There exist a set of functions $\{\sigma_{\alpha}(u)\}$
of class C$^{\infty}$ which satisfy
\begin{align}
\sigma_{\alpha}(u)\geq0,  &  \quad\sum_{\alpha}\sigma_{\alpha}%
(u)=1,\nonumber\\
\sigma_{\alpha}(u)>0\;(u\in\lbrack0,b)^{d}),  &  \quad\mathrm{supp}\text{
}\sigma_{\alpha}(u)=[0,b]^{d},\nonumber
\end{align}
such that, for arbitrary integrable function $H(\theta)$,
\begin{align*}
\int_{\Theta_{\epsilon}}H(\theta)\pi(\theta)d\theta &  =\int_{\mathcal{U}%
}H(g(u))\pi(g(u))|g^{\prime}(u)|du\\
&  =\sum_{\alpha}\int_{U_{\alpha}}H(g(u))\tilde{\pi}^{\ast}(u)u^{h}du,
\end{align*}
where we define $\tilde{\pi}^{\ast}(u)$ by omitting local coordinate $\alpha
$,
\[
\tilde{\pi}^{\ast}(u)=\sigma_{\alpha}(u)\tilde{\pi}(u).
\]

\end{theo}

\begin{proof}
See \cite{W1}.
\end{proof}

\begin{defi}%
\begin{align}
Z_{n}=\int\prod_{j=1}^{n}p(\rho_{j}|\theta)^{\beta}\pi(\theta)d\theta &
,\qquad Z_{n}^{0}=\frac{Z_{n}}{\displaystyle{\prod_{j=1}^{n}p_{0}(\rho
_{j})^{\beta}}}\\
F_{n}=-\frac{1}{\beta}\log Z_{n}  &  ,\qquad F_{n}^{0}=-\frac{1}{\beta}\log
Z_{n}^{0}%
\end{align}
$Z_{n}$ and $F_{n}$ are called a partition function and a Bayes stochastic
complexity, respectively.
\end{defi}

The zeta function $\displaystyle{\zeta(z)=\int D(\theta)^{z}\pi(\theta
)d\theta}$ can be analytically continued to the unique meromorphic function on
the entire complex plane. All poles of $\zeta(z)$ are real, negative, rational
numbers.
\begin{align}
(-\lambda)  &  :=\mathrm{maximum}\;\mathrm{pole}\;\mathrm{of}\;\zeta
(z)\quad(\lambda>0),\\
m  &  :=\mathrm{multiplicity}\;\mathrm{of}\;(-\lambda).
\end{align}
$\lambda$ and $m$ are called the learning coefficient and its order,
respectively. If $D(\theta)$ and a priori distribution $\pi(\theta)$ are
represented in Theorem 2.8 and 2.9, then the learning coefficient and its
order are given, respectively, by
\begin{align}
\lambda &  =\min_{\alpha}\min_{1\leq j\leq d}\left(  \frac{h_{j}+1}{2k_{j}%
}\right)  ,\\
m  &  =\max_{\alpha}\#\{j\;|\;\lambda=(h_{j}+1)/2k_{j}\},
\end{align}
where \# denotes the cardinality of the set. Let $\{\alpha^{\ast}\}$ be a set of all local
coordinates in which both the minimization in Eq.(45) and the maximization in
Eq.(46) are attained. Such a set of local coodinates $\{\alpha^{\ast}\}$ is
said to be the essential family of local coordinates. For each local
coordinate $\alpha^{\ast}$ in the essential family of local coordinates, we
assume without loss of generality $u$ is represented as $u=(x,y)$ so that
\begin{align*}
x  &  =(u_{1},u_{2},\cdots,u_{m}),\\
y  &  =(u_{m+1},u_{m+2},\cdots,u_{d}),
\end{align*}
and that
\begin{align*}
\lambda &  =\frac{h_{j}+1}{2k_{j}}\;\;(1\leq j\leq m),\\
\lambda &  <\frac{h_{j}+1}{2k_{j}}\;\;(m+1\leq j\leq d).
\end{align*}
For any function $H(u)=H(x,y)$, we use the notation $H_{0}(y):=H(0,y)$.

\begin{theo}
\label{asymptoticbehavior}$(1)$
\begin{align}
F_{n}^{0}-\frac{\lambda}{\beta}\log n+\frac{m-1}{\beta}  &  \log\log
n\nonumber\\
\longrightarrow-\frac{1}{\beta}\log\left(  \sum_{\alpha^{\ast}}\right.   &
\gamma_{b}\left.  \int_{0}^{\infty}dt\int_{U_{\alpha^{\ast}}}t^{\lambda
-1}e^{-\beta t+\beta\sqrt{t}\xi_{0}(y)}\tilde{\pi}_{0}^{\ast}(y)dy\right)
\;\;\;in\;law.
\end{align}
$(2)$ The following asymptotic expansion holds:
\begin{equation}
F_{n}=nL_{n}+\frac{\lambda}{\beta}\log n-\frac{m-1}{\beta}\log\log n+F_{n}%
^{R},
\end{equation}
where $\displaystyle{L_{n}=-\frac{1}{n}\sum_{j=1}^{n}\log p_{0}(\rho_{j})}$,
and $F_{n}^{R}$ is a random variable which converges in law to a random variable.
\end{theo}

\begin{proof}
We can prove (1) easily by using the same method as used in \cite{W1}. We
define
\begin{equation}
F_{n}^{R}=-\frac{1}{\beta}\log\left(  \sum_{\alpha^{\ast}}\right.  \gamma
_{b}\left.  \int_{0}^{\infty}dt\int_{U_{\alpha^{\ast}}}t^{\lambda-1}e^{-\beta
t+\beta\sqrt{t}\xi_{n,0}(y)}\tilde{\pi}_{0}^{\ast}(y)dy\right)  .
\end{equation}
Then, (2) immediately derives from (1).
\end{proof}

This theorem clarifies the behaviour of the Bayes likelihood function which
evaluates how close a model approaches to an optimal state, according to the
increase of data. Although there is no essential difference between this
theorem and that in classical case \cite{W1}, there is one reason why we
described this result here: The results such as Theorem \ref{standardform},
\ref{asymptoticbehavior}, and the next Theorem \ref{WAIC} are the standard
objects of interest in modern statistical science, and fundamental analysis in
information theory is mainly based on large-deviation type results similar to
Theorem \ref{asymptoticbehavior}. Therefore, this therem is of vital
importance and need to be investigated in more details regardless of quantum
or classical in future.

For a given function $G(\theta)$ on $\Theta$, the a posteriori mean of
$G(\theta)$ is defined as
\begin{equation}
\langle G(\theta)\rangle_{\pi,\beta}^{\rho^{n}}=\frac{\displaystyle{\int
G(\theta)\prod_{j=1}^{n}p(\rho_{j}|\theta)^{\beta}\pi(\theta)d\theta}%
}{\displaystyle{\int\prod_{j=1}^{n}p(\rho_{j}|\theta)^{\beta}\pi
(\theta)d\theta}},
\end{equation}
where $0<\beta<\infty$. Then, the following equality holds:
\begin{equation}
\omega_{\pi,\beta}^{n}=\int\rho\;\langle p(\rho|\theta)\rangle_{\pi,\beta
}^{\rho^{n}}\;dm(\rho).
\end{equation}

\begin{defi}
$\;$\newline$(1)$ Bayes generalization error and Bayes generalization loss are
defined, respectively, by
\begin{equation}
\mathcal{E}_{bg}=E_{\rho}\left[  \log\frac{q(\rho)}{\langle p(\rho
|\theta)\rangle_{\pi,\beta}^{\rho^{n}}}\right]  ,\qquad\mathcal{L}%
_{bg}=E_{\rho}\left[  -\log\langle p(\rho|\theta)\rangle_{\pi,\beta}^{\rho
^{n}}\right]  .
\end{equation}
$(2)$ Bayes training error and Bayes training loss are defined, respectively,
by
\begin{equation}
\mathcal{E}_{bt}=\frac{1}{n}\sum_{j=1}^{n}\left[  \log\frac{q(\rho_{j}%
)}{\langle p(\rho_{j}|\theta)\rangle_{\pi,\beta}^{\rho^{n}}}\right]
,\qquad\mathcal{L}_{bt}=\frac{1}{n}\sum_{j=1}^{n}\left[  -\log\langle
p(\rho_{j}|\theta)\rangle_{\pi,\beta}^{\rho^{n}}\right]  .
\end{equation}
\newline$(3)$ functional variance is defined by:
\begin{equation}
\mathcal{V}=\sum_{j=1}^{n}\left\{  \langle(\log p(\rho_{j}|\theta))^{2}%
\rangle_{\pi,\beta}^{\rho^{n}}-(\langle\log p(\rho_{j}|\theta)\rangle
_{\pi,\beta}^{\rho^{n}})^{2}\right\}  .
\end{equation}

\end{defi}

These notions are the main targets to be estimated or calculated in statistics
and learning theory. We can easily check that
\begin{align}
\mathcal{E}_{bg}  &  =D(q\Vert\langle p(\cdot|\theta)\rangle_{\pi,\beta}^{\rho^{n}})
=S(\psi\Vert\omega_{\pi,\beta}^{n}),\\
\mathcal{E}_{bg}  &  =\mathcal{L}_{bg}+E_{\rho}\left[  \log q(\rho
_{j})\right]  ,\nonumber\\
\mathcal{E}_{bt}  &  =\mathcal{L}_{bt}+\frac{1}{n}\sum_{j=1}^{n} \log
q(\rho_{j}).\nonumber
\end{align}
Our present concern is the following theorem.

\begin{theo}
\label{WAIC}
\begin{align}
E[\mathcal{L}_{bg}]  &  =E[\mathrm{WAIC}]+o\left(  \frac{1}{n} \right)  ,\\
\mathrm{WAIC}  &  =\mathcal{L}_{bt} +\frac{\beta}{n}\mathcal{V}.
\end{align}

\end{theo}

\begin{proof}
See \cite{W3}.
\end{proof}

WAIC is the acronym for \textquotedblleft widely applicable information
criteria\textquotedblright. It is shown by this theorem that the WAIC for a central measure is asymptotically equal
to the Bayes generalization loss. Since WAIC for $p_{\theta}=\dfrac
{d\mu_{\omega_{\theta}}}{dm}$ is a quantum version of the information criteria
(IC), this result can be successfully interpreted as establishing IC for
quantum states. This also justifies our use of the central measure
$\mu_{\omega}$ for the central decomposition of $\omega\in E_{\mathfrak{A}}$:
namely, owing to the use of central decomposition, our LDS in the second level
can determine representations controlling spectra of observables, on the basis
of numerical data of such a quantity as WAIC. It is important, not only
practically but also conceptually, that such qualitative aspects as
representations of the algebra of observables can be estimated by this kind of
quantitative data. In addition, WAIC in quantum case should be contrasted with
that in classical case, since the latter cannot evaluate representations of the
algebra. On the other hand, we note that IC in the first level are the same as
those in classical case.

\subsubsection{Physical meaning and practical use}

We can conclude that we have established the following procedures:

\begin{center}
Rate function$\;\Rightarrow$ Predictive state$\;\Rightarrow$ Information
criterion ($\Rightarrow$ \textquotedblleft True\textquotedblright\ state).
\end{center}

\noindent The procedure established in Section 2.3.1 is a typical example of this:

\begin{center}
Quantum relative entropy $S(\cdot\Vert\cdot)$ $\;\Rightarrow$\newline Bayesian
escort predictive state $\omega_{\pi,\beta}^{n}$ $\;\Rightarrow$
$\displaystyle{\mathrm{WAIC}=\mathcal{L}_{bt}+\frac{\beta}{n}\mathcal{V}}$
($\Rightarrow$ \textquotedblleft True\textquotedblright\ state $\psi$).
\end{center}

First, rate functions are specified by procedures in LDP. A rate function is a
barometer to what extent one state diverges from a \textquotedblleft
true\textquotedblright$\;$one. Secondly, we construct predictive states from
models and data by applying the results of several steps whose starting point
is the rate function provided by the first step. Thirdly, we define IC and use
it for selecting the best predictive state from candidates. Lastly, we select
one state which should be treated as a \textquotedblleft true\textquotedblright$\;$one
$\psi$ in Section 2.3.1.
Taking this step, we can reach a \textquotedblleft true\textquotedblright$\;$state by
using the methods in Section 2.3.1 such as Theorem \ref{standardform},
\ref{asymptoticbehavior}, and \ref{WAIC}. As stated in Section 2.3.1, IC are
estimators for rate functions as quasi-distances from a \textquotedblleft
true\textquotedblright$\;$state to a predictive state, which have bias terms
based on the method to construct predictive states.

\section{Examples}

Once the sector structure consisting of mutually disjoint factor states is
clarified, the Micro-Macro duality starts to be valid, according to which the
present method of LDS becomes effective. From this viewpoint, the following
examples are instructive in the sense that the method in LDS second level
enable us to reduce complicated dynamical systems partially to
kinematics.\newline(1) Non-equilibrium states in quantum field theory

The method established in \cite{BOR02} is used for describing non-equiliburium
states in QFT. The universal model of the relevant sector structure to this
context is known to be provided by a family of factor KMS
(Kubo-Martin-Schwinger) states $\{\omega_{\beta,\mu}|\beta>0,\mu\in K\}$ on a
von Neumann algebra $\mathfrak{M}$ of type III parametrized by the inverse
temperature $\beta$ and by all other necessary thermodynamic parameters
denoted collectively by $\mu\in K$ such as a chemical potential. Following the
ideas in \cite{BOR02}, we can write a non-equilibrium state of the system
whose reference states are $\{\omega_{\beta,\mu}|\beta>0,\mu\in K\}$ on a von
Neumann algebra $\mathfrak{M}$ of type III as follows:
\begin{equation}
\omega_{B,\rho}=\int_{B}d\rho(\beta,\mu)\;\omega_{\beta,\mu},
\end{equation}
where $B$ is a compact subset of $\mathbb{R}_{>0}\times K$ and $\rho$ is a
regular Borel measure on $\mathbb{R}_{>0}\times K$. In this situation, we can
construct a model $\{\rho_{\theta}(\beta,\mu)|\theta\in\Theta\subset
\mathbb{R}^{d}:\mathrm{compact}\}$ of probability distribution, in terms of
which the method of statistical inference can be systematically applied for
the purpose of further developments of the theory of non-equilibrium states in
QFT. \newline(2) Conformal field theory and critical phenomena

Let $\mathfrak{C}$ be a C$^{\ast}$-algebra generated by $\{e^{iL_{n}%
},e^{iC}|n\in\mathbb{Z}\}$ such that operators $\{L_{n}\}$ and a self-adjoint
operator $C$ on a Hilbert space $\mathfrak{H}$ satisfy
\begin{equation}
\lbrack L_{m},L_{n}]=(m-n)L_{m+n}+\frac{m^{3}-m}{12}C\delta_{m+n,0}%
,\;[L_{n},C]=0.
\end{equation}
Let $\{\omega_{c}|c\in Spec(C)\subset\mathbb{R}\}$.
\begin{equation}
\omega_{R}=\int_{R}d\sigma(c)\;\omega_{c},
\end{equation}
where $\sigma$ is a regular Borel measure on $Spec(C)$ and $R$ is compact
subset of $Spec(C)$. In view of the accumulated applications of conformal
field theory, it would be natural to expect the possibility of systematic
theory for statistical estimate about critical phenomena in solid state
physics on the basis of the mathematical knowledge about the reducible
representations and states of this kind, which would be the target for future tasks.

\section{Discussion on Quantum Estimation Theory: Quantum Model Selection}

In Section 2.3.2, we have discussed estimation theory for quantum states.
Remarkably, the methods developed here allow us to take full advantage of the
usual measure-theoretical analysis in statistics, information theory and
learning theory even for estimation of quantum states in the context of
quantum theory, which is due to our bringing the use of the central measure
$\mu_{\omega}$ of $\omega\in E_{\mathfrak{A}}$ into focus. This should be
contrasted with many previous attempts in quantum estimation theory, where
vain efforts have been expended for the attempts of formulating new notions or
\textquotedblleft quantized version\textquotedblright\ of the notions known in
classical (measure-theoretical) statistics, information theory and learning
theory. Instead, what is most crucial here is the difference in the method of
inference according to whether a state to be estimated is factor or not. Since
the methods discussed in Section 2.3 are for non-factor states, different
analysis from the one for factor states need to be built up. On the other hand, 
we have succeeded in constructing quantum model selection, which is a quantum version
of model selection, by using measure-theoretic methods. Model selection began
when Akaike introduced the concept of information criteria in 1971
\cite{Ak73,Ak74} to resolve the insufficiency of hypothesis testing for
selecting the best predictive distribution. The best known and used one is the
Akaike information criterion (AIC)
\begin{equation}
\mathrm{AIC}=-\frac{1}{n}\sum_{j=1}^{n}p(x_{j}|\hat{\theta}_{MLE})+\frac{d}{n}%
\end{equation}
where $\hat{\theta}_{MLE}$ is the maximal likelihood estimator (MLE) and $d$
is the dimension of parameters. AIC can be applied in the situation that the
maximal likelihood method, or the M-estimation method, is used for regular
models. Furthermore, WAIC appearing in singular statistics \cite{W1,W2} is
another version of IC and contains AIC and TIC as a special case. Roughly
speaking, model selection is the method for selecting the
predictive distribution which attains the minimum of IC in several candidates.
Although AIC has been used for quantum states in \cite{UNTMN03,YE11}, the reason
has not been clarified why we can apply it for quantum states. Because they
applied AIC to a general positive operator-valued measure (POVM), not to
PVM's, their use is not precisely in the first level. However, with the help of
measuring processes and Naimark dilation, we can justify their use. Thus it is
desirable to examine the validity of the use of IC for quantum states.

\begin{rema}
It is occasionally said that, by using AIC, or BIC, some model with fewer
parameters is automatically chosen. However, this statement is not
precise and is no more than hindsight: if two models have almost equal
training errors, then AIC of the model with fewer parameters becomes smaller
than the others, and the model having the smallest AIC is naturally chosen. As
stated in Section 2.3.2, we should fix a \textquotedblleft
true\textquotedblright$\;$state by using the predictive state and test the
performance of the latter compared with other predictive states. Therefore, we
should use flexibly the predictive state selected by IC without taking it by
absolute priority.
\end{rema}

In recent years, algebraic geometry and algebraic analysis are successfully
applied to the singular aspects in learning theory. Many statistical models,
such as the normal mixture model
\begin{equation}
f(x|\mathbf{a},\mathbf{b},\mathbf{c})=\sum_{j=1}^{M}a_{j}\varphi(x|b_{j}%
,c_{j}),
\end{equation}
where $\mathbf{a}=(a_{1},\cdots,a_{M})$ such that $a_{1},\cdots,a_{M}\geq0$
and $\displaystyle{\sum_{j=1}^{M}a_{j}=1}$, $\mathbf{b}=(b_{1},\cdots
,b_{M})\in\mathbb{R}^{M}$, $\mathbf{c}=(c_{1},\cdots,c_{M})\in(\mathbb{R}%
_{+})^{M}$ and $\displaystyle{\varphi(x|b,c)=\frac{1}{(2\pi c^{2})^{1/2}}%
\exp\left\{  -\frac{1}{2c^{2}}(x-b)^{2}\right\}  }$, have degenerate Fisher
information matrices, so that Riemannian-geometric methods cannot be applied.
Then the Cram$\mathrm{\acute{e}}$r-Rao inequality
\begin{equation}
V(\theta)\geq J^{-1}(\theta),
\end{equation}
does not hold without any significance, where $V(\theta)=(E_{x}[(\hat{\theta
}_{j}(x)-\theta_{i})(\hat{\theta}_{j}(x)-\theta_{j})])_{i,j}$ and
$J^{-1}(\theta)$ are, respectively. the covariance matrix for $\theta=(\text{\boldmath$a$%
},\text{\boldmath$b$},\text{\boldmath$c$})$ and the inverse of the Fisher
information matrix $J(\theta)$. Therefore, different methods
using algebraic geometry and algebraic analysis are investigated, which work
efficiently in various areas, and are already used in a textbook
\cite{DSS08}, which has encouraged us to use algebraic geometric methods.

Lastly, we give a generalization of quantum hypothesis testing. Suppose that
$\varphi,\psi\in E_{\mathfrak{A}}$ have central measures given, for any integrable function $f$, by
\begin{align*}
\int f(\rho) d\mu_{\varphi}(\rho)  &  =\sum_{j=1}^{m}\alpha_{j}f(\rho_{j}),\\
\int f(\rho) d\mu_{\psi}(\rho)  &  =\sum_{j=1}^{m}\beta_{j}f(\rho_{j}),
\end{align*}
with $\sum_{j}\alpha_{j}=\sum_{j}\beta_{j}=1$, $0<\alpha_{j},\beta_{j}<1$
$(j=1,\cdots,m)$, corresponding, respectively, to the following diagonal
matrices:
\begin{equation}
\mu_{\varphi}\leftrightarrow\sigma_{\varphi}=\left(
\begin{array}
[c]{cccc}%
\alpha_{1} &  &  & O\\
& \alpha_{2} &  & \\
&  & \ddots & \\
O &  &  & \alpha_{m}%
\end{array}
\right)  ,\;\mu_{\psi}\leftrightarrow\sigma_{\psi}=\left(
\begin{array}
[c]{cccc}%
\beta_{1} &  &  & O\\
& \beta_{2} &  & \\
&  & \ddots & \\
O &  &  & \beta_{m}%
\end{array}
\right)  .
\end{equation}
Let $E_{\psi}(\Delta)$ $(\Delta\in\mathcal{B}(\mathrm{supp}\;\mu_{\psi}))$ be
the PVM corresponding to $\mu_{\psi}$ (see Eq.(\ref{centralPVM})). It is
immediately seen that $S(\varphi\Vert\psi)=D(\mu_{\varphi}\Vert\mu_{\psi
})=S(\sigma_{\varphi}\Vert\sigma_{\psi})$. We treat the state $\psi$ as a
\textquotedblleft true\textquotedblright\ one and assume that a test function of
interest $S^{n}:(\mathrm{supp}\;\mu_{\psi})^{n}\mapsto\{0,1\}$ 
$S^{n}$ has a positive
operator representation $A_{n}$ on $M(m,\mathbb{C})^{\otimes n}$ such that
$0\leq A_{n}\leq I_{M(m,\mathbb{C})^{\otimes n}}$. Then, the error
probabilities of the first kind and the second kind can, respectively, be
defined by
\begin{align*}
\alpha_{n}(A_{n})  &  =\mathrm{Tr}[\sigma_{\psi}^{\otimes n}(I-A_{n})],\\
\beta_{n}(A_{n})  &  =\mathrm{Tr}[\sigma_{\varphi}^{\otimes n}A_{n}].
\end{align*}
The following theorem is a generalization of quantum Stein's lemma and can be
proved by the same method as found in \cite{HP91,ON00,Asym05}, valid for
the version of quantum Stein's theorem in these papers.

\begin{theo}
\label{Stein} For any $0<\epsilon<1$, it holds that
\begin{equation}
\lim_{n\rightarrow\infty}\frac{1}{n}\log\beta_{n}^{\ast}(\epsilon
)=-S(\psi\Vert\varphi),
\end{equation}
where $\beta_{n}^{\ast}(\epsilon)$ is the minimum second error probability
under the constraint that the first error probability is less than $\epsilon$,
i.e.,
\begin{equation}
\beta_{n}^{\ast}(\epsilon)=\{\beta_{n}(A_{n})\vert A_{n}\in M(m,\mathbb{C}%
)^{\otimes n} ,0\leq A_{n}\leq I_{M(m,\mathbb{C})^{\otimes n}},\alpha
_{n}(A_{n})\leq\epsilon\}.
\end{equation}

\end{theo}

It is important that $\{A_{n}\}$ are merely operator-valued representations of
tests $\{S_{n}\}$ without actual uses in measurements, where $E_{\psi}(\Delta)$ is
actually used. It is obvious that the quantum relative entropy formulated in the 
present paper is accessible to actual experimental situations whose operational 
meaning is different from that in \cite{Asym05,HP91,OH04,ON00} formulated in quantum
i.i.d. states.

\section{Conclusion and Perspective}

In this paper we have proposed Large Deviation Strategy and established its
first and second levels. For this purpose, we have clarified that the quantum
relative entropy plays the role of the rate function in LDP 2nd level, according 
to which several measure-theoretical methods work efficiently in quantum case. 

While results of this sort have been anticipated on the basis of a simple 
analogy to the classical case or direct computations in some special situations, 
the pertinence of such formal derivations has been questionable for lack of 
the appropriate operational setting-up to guarantee the appropriate interpretation. 
In the present case, we can safely use the natural relation, 
$S(\psi\Vert\omega)=D(\nu\Vert\mu)$, due to \cite{HOT83} to bridge the quantum
context with the classical one, which sweeps away all the suspicions.

However, the situation about the estimation theory for the internal structures
of a factor state is quite different, which seems to require some new ideas.
For this purpose, the measurement scheme formulated in \cite{Oj05, HO09} would
be instructive as its aim is to search the internal structure of a factor
state. To proceed further along the present line of thoughts, the tasks to
construct estimation theory for factor states and to establish the third and
fourth levels in LDS will be crucially important.

\subsection*{Acknowledgements}

The authors would like to thank Messrs. Hayato Saigo, Ryo Harada, Takahiro
Hasebe, and Hiroshi Ando for helpful discussions and comments. One of them
(K.O.) is grateful to Prof. Tatsuaki Wada and Dr. Fuyuhiko Tanaka for
discussions and their interests, comments and encouragements. He would like to 
express deep gratitude to Prof. Takashi Kumagai for valuable instructions and
encouragements. They also thank the Yukawa Institute for Theoretical Physics
at Kyoto University, where this work was reported during the YITP-W-10-14 on
\textquotedblleft Duality and Scales in Quantum-Theoretical
Sciences\textquotedblright.

\appendix

\section{Barycentric Decomposition of States and Extension of Algebra}

In the barycentric decomposition, $\displaystyle{\omega=\int_{E_{\mathfrak{A}%
}}\rho\; d\mu(\rho)}$, of a state $\omega$ of $\mathfrak{A}$ by an orthogonal
measure $\mu$ we have a spectral measure $E_{\mu}:=(\mathcal{B}(\mathrm{supp}%
\;\mu)\ni\Delta\mapsto E_{\mu}(\Delta):=\kappa_{\mu}(\chi_{\Delta}%
)\in\mathfrak{B})$ on $E_{\mathfrak{A}}$, taking values in a subalgebra
$\mathfrak{B}$ of the commutant $\pi_{\omega}(\mathfrak{A})^{\prime}$ but not
in $\pi_{\omega}(\mathfrak{A})^{\prime\prime}$ of observables. When we
consider a physical process described by this spectral measure, it involves
the object system with $\pi_{\omega}(\mathfrak{A})^{\prime\prime}(\subset
L^{\infty}(E_{\mathfrak{A}},\mu))$ and the measuring one $\mathfrak{B}%
(\subset\pi_{\omega}(\mathfrak{A})^{\prime})$, the latter of which registers
the indices to determine states of $\mathfrak{A}$. According to the
measurement scheme \cite{Oj05, HO09}, we have a composite system consisting of
these two algebras $\pi_{\omega}(\mathfrak{A})^{\prime\prime}$ and
$\mathfrak{B}$ through a suitable measurement coupling, which amounts to the
extension of algebra:
\begin{equation}
\pi_{\omega}(\mathfrak{A})^{\prime\prime}\hookrightarrow\pi_{\omega
}(\mathfrak{A})^{\prime\prime}\vee\mathfrak{B}\cong\pi_{\omega}(\mathfrak{A}%
)^{\prime\prime}\otimes\mathfrak{B}\subset\pi_{\omega}(\mathfrak{A}%
)^{\prime\prime}\vee\pi_{\omega}(\mathfrak{A})^{\prime}=\mathfrak{Z}%
_{\pi_{\omega}}(\mathfrak{A})^{\prime}.
\end{equation}
In this context, we can extend the state $\omega\in E_{\mathfrak{A}}$ to the
one $\tilde{\omega}$ on the algebra $\pi_{\omega}(\mathfrak{A})^{\prime\prime
}\vee\mathfrak{B}$ simply defined for $A\in\pi_{\omega}(\mathfrak{A}%
)^{\prime\prime}\vee\mathfrak{B}$ by
\begin{equation}
\tilde{\omega}(A)=\langle\Omega_{\omega},A\Omega_{\omega}\rangle.
\end{equation}
The naturality of this procedure is understood in relation with
Tomita-Takesaki modular theory \cite{BR1,T02}, which plays vital roles in LDS
2nd level to define the quantum relative entropy. Here we note also that
$\mathfrak{B}$ is an abelian subalgebra of $\pi_{\omega}(\mathfrak{A}%
)^{\prime}$ and that $\pi_{\omega}(\mathfrak{A})^{\prime}=J\pi_{\omega
}(\mathfrak{A})^{\prime\prime}J$ in terms of the modular conjugation operator
$J$ which lives in the standard representation of $\pi_{\omega}(\mathfrak{A})^{\prime\prime}$. 
The most important barycentric measures apart from central ones are
extremal measures corresponding to a maximal abelian subalgebra of
$\pi_{\omega}(\mathfrak{A})^{\prime}$. The measure $\mu$ is pseudosupported by
the pure states $\mathcal{E}(E_{\mathfrak{A}})$ over $\mathfrak{A}$.

Let $\mu$ be an orthogonal measure with a barycenter $\psi\in E_{\mathfrak{A}%
}$ such that there is a subcentral measure $m$ satisfying $\mu\ll m$. By the
same discussion as in Section 2.3.1, we define for any $\Gamma\in
\mathcal{B}_{cy}(M_{1}(E_{\mathfrak{A}}))$,%
\begin{equation}
Q_{n}^{(2A)}(\Gamma)=P_{\mu}(L_{n}\in\Gamma).
\end{equation}
Then, the next theorem holds.

\begin{theo}
Let $\mathfrak{A}$ be a separable C$^{\text{*}}$-algebra, $\psi$ be a state on
$\mathfrak{A}$, and $\mu$ be a barycentric measure of $\psi$. Then
$Q_{n}^{(2A)}$ satisfies LDP with the rate function $D(\cdot\Vert
\mu)(=S(b(\cdot)\Vert\psi))$:
\begin{align}
-\underline{D}(\Gamma\Vert\mu):=-\inf_{\nu\in\Gamma^{o}}D(\nu\Vert\mu)  &
\leq\liminf_{n\rightarrow\infty}\frac{1}{n}\log Q_{n}^{(2A)}(\Gamma
)\nonumber\\
\leq\limsup_{n\rightarrow\infty}\frac{1}{n}\log Q_{n}^{(2A)}(\Gamma)  &
\leq-\inf_{\nu\in\overline{\Gamma}}D(\nu\Vert\mu)=:-\overline{D}(\Gamma
\Vert\mu)
\end{align}
for any $\Gamma\in\mathcal{B}_{cy}(M_{1}(E_{\mathfrak{A}}))$. If there
exists a subcentral measure $m$ such that $\nu,\mu\ll m$, then $D(\nu\Vert
\mu)=S(b(\nu)\Vert\psi)$ holds for such $\nu$ belonging to $\overline{\Gamma}$
(or $\Gamma^{o}$) that $D(\nu\Vert\mu)=\overline{D}(\Gamma\Vert\mu)$ or $\underline{D}(\Gamma\Vert\mu)$.
\end{theo}

The results in Section 2.3.1 and Section 4 also hold for general barycentric
measures. It is, however, necessary to keep in mind that barycentric
decompositions in general do not always have clear physical meaning, in
contrast to a central decomposition. To extend quantum estimation theory this
point has to be resolved.

\end{document}